\documentclass[12pt]{article}
\newcommand{\be}{\begin{equation}}
\newcommand{\ee}{\end{equation}}
\newcommand{\bea}{\begin{eqnarray}}
\newcommand{\eea}{\end{eqnarray}}
\newcommand{\sptwo}{1.4}
\newcommand{\doublespace}{\edef\baselinestretch{\sptwo}\Large\normalsize}
\newcommand{\newsection}[1]{\setcounter{equation}{0}}

\newcounter{newapp}
\begin{document}
\begin{titlepage}
\vspace*{1.0in}
\begin{center}
{\large\bf Incompatibility Of Simultaneous Non-Linear Realizations \\
Of Scale Symmetry and Supersymmetry }
\end{center}
\vspace{0.25in}
~\\
\begin{center}
{\bf T.E. Clark and S.T. Love}\\
{\it Department of Physics\\ 
Purdue University\\
West Lafayette, IN 47907-1396}
~\\
~\\
\end{center}
\vspace{1in}
\begin{center}
{\bf Abstract}
\end{center}
Simultaneous nonlinear realizations of supersymmetry and softly broken
scale and chiral symmetries are investigated. To guarantee
Nambu-Goldstone realizations of the symmetries, the Goldstino decay
constant is forced to vary as the explicit soft scale and chiral
symmetry breaking parameters. Consequently, it must vanish in the chiral
limit and the simultaneous nonlinear realizations of the super and scale
symmetries proves inconsistent.   
\end{titlepage}
\pagebreak

\doublespace

\newsection

Goldstone's theorem \cite{G} guarantees that associated with every
spontaneously broken global symmetry there is a massless particle. Below
the symmetry breaking scale, the dynamics of these Nambu-Goldstone
degrees of freedom can be descibed by an action which realizes the
spontaneously broken symmetry nonlinearly. This effective action
encapsulates all the consequences of the symmetry current algebra and,
through lowest non-trivial order in a derivative expansion, is unique up
to reparametrization, independent of the underlying theory
\cite{SW}-\cite{CCWZ}. 

For the case of an internal global symmetry group $G$ spontaneously
broken to an invariant subgroup $H$, the Nambu-Goldstone fields,
$\pi^i~,~i=1,...{\rm dim} G/H$, act as coordinates of the coset manifold
$G/H$. A particular choice of the coset coordinates is the standard
realization parametrized as 
\be
U(\pi)=e^{\frac{2iT^i \pi^i}{F_\pi}} ,
\ee
where $T^A$ is the fundamental representation of $G$ and $F_\pi$ is the
Nambu-Goldstone boson decay constant. While $U$ transforms linearly
under $G$, the Nambu-Goldstone fields, $\pi^i$, transform linearly only
under $H$ and nonlinearly under the spontaneously broken $G$ generators.
With  this choice of coordinates, the $G$ invariant effective Lagrangian
is simply \cite{HG}
\be
{\cal L}=\frac{F_\pi^2}{4}{\rm Tr}\left[ \partial_\mu U^\dagger
\partial^\mu U\right].
\ee
In addition to the spontaneous symmetry breaking, there also often
appears some soft explicit symmetry breaking. For the case of chiral
symmetry breaking, this explicit breaking takes the form of a soft mass
term and the above effective Lagrangian is modified  to
\be
{\cal L} = \frac{F_\pi^2}{4}{\rm Tr}\left[ \partial_\mu U^\dagger
\partial^\mu U\right] -u F_\pi^2 {\rm Tr}\left[mU^\dagger +Um\right] .
\label{GB}
\ee
Here $m$ is the mass matrix characterizing the soft explicit breaking
and $u$ is the order parameter of the spontaneous symmetry breaking. For
example, if the chiral symmetry is dynamically broken due to some
underlying strong gauge interaction, then 
$u=\frac{<\bar\psi \psi>}{2F_\pi^2}$, where $\psi$ is a chiral fermion
of the underlying theory. The above effective Lagrangian assumes that
the theory is free of chiral anomalies. If such effects are also
present, the effective action is further modified by the inclusion of a
Wess-Zumino term \cite{WZ}-\cite{CL}. While we have focused on the case
of a single spontaneously broken global symmetry, the analysis is
trivially extended to include the possibility of multiple spontaneously
and softly broken internal global symmetry groups. The resultant
effective Lagrangian is simply obtained by additively including the
individual effective Lagrangians. 

In addition to the case of spontaneously broken global symmetries, one
can also construct effective Lagrangians which nonlinearly realize scale
symmetry. Thus we envision an underlying model where over some energy
range the quantum fluctuations are such that the scale anomaly either
vanishes or is but a very small effect and dilatations are approximately
a good symmetry broken only by some soft mass terms. Then when a global
internal symmetry is spontaneously broken, there will be an accompanying
spontaneous scale symmetry breakdown \cite{BLL1}-\cite{BL} and the
spectrum will include its associated Nambu-Goldstone boson, the dilaton
$D$. Since, in general, the couplings of the underlying theory do indeed
run, the dilaton is a pseudo Nambu-Goldstone boson aquiring a mass
related to the scale at which the renormalization group $\beta$
functions become significant. To nonlinearly realize the scale symmetry,
one introduces a standard realization for the dilaton as
\be
S(D)=e^{\frac{D}{F_D}}
\ee
where $F_D$ is the dilaton decay constant. The associated scale
transformation, \\
parametrized by $\epsilon$, is 
\be
\delta^D (\epsilon)S=\epsilon(1+x^\nu\partial_\nu)S 
\ee
and results in an inhomogeneous scale transformation of the dilaton as 
\be
\delta^D (\epsilon)D=\epsilon(F_D + x^\nu \partial_\nu D) .
\ee
Since the generators of space-time scale transformations commute with
those of the internal symmetry group $G$, the Nambu-Goldstone boson
fields $\pi^i$ are constrained \cite{CLL} to carry zero scale weight
\be
\delta^D (\epsilon)\pi^i =  \epsilon x^\nu \partial_\nu \pi^i ,
\ee
while the dilaton is a $G$ singlet and thus satisfies
\be
\delta^G(\omega) D =0 ,
\ee
with $\omega^A$ parametrizing the group $G$ transformations.

The effective action (\ref{GB}) can be made scale invariant up to soft
breaking terms by including appropriate powers of $S$ to make the
scaling weight of each invariant term be four. The soft explicit scale
and $G$ symmetry breaking terms are dictated by the form of the
underlying theory. For the case of a global chiral symmetry broken by a
soft fermion mass term which also softly breaks the scale symmetry, the
effective Lagrangian which nonlinearly realizes both the chiral and the
scale symmetry takes the form
\cite{WB}-\cite{BLL3}
\bea
 {\cal{L}} &=& \frac{1}{2} F_D^2 \partial_\mu S \partial^\mu S 
                + \frac{1}{4} F_\pi^2 S^2 Tr[\partial_\mu U^\dagger
                  \partial^\mu U]\\
             &~& -\frac{1}{2} u F^2_\pi (3-\gamma)S^4 Tr[m] 
+ u F_\pi^2 S^{3-\gamma} Tr[mU^\dagger  + U m].  
\label{SC}
\eea
where $\gamma$ is the anomalous dimension of the underlying fermion
field. 
Under the nonlinear scale and chiral symmetries, this effective
Lagrangian transforms as 
\be
\delta^D (\epsilon) {\cal{L}} = \epsilon (3+x^\nu \partial_\nu)\{u
F^2_\pi (3-\gamma) S^{3-\gamma} Tr [mU^\dagger + U m]\}
\ee
\be
\delta^G (\omega) {\cal{L}} = -i\omega^A u F^2_\pi S^{3-\gamma}Tr[T^A (m
U^\dagger - U m)]
\ee
where the right hand sides reflect the soft explicit symmetry breaking.

The fact that the coefficient of the scale and chirally invariant
$S^4$ term in the Lagrangian (\ref{SC}) depends on the explicit
breaking mass parameter, $m$, is somewhat unusual and 
warrants some further elaboration. The necessity for this value can be
established by expanding $\cal L$ in powers of the dilaton
field $D$.  The elimination of the destabilizing term linear in $D$ is 
accomplished by fixing it to be  
$-\frac{1}{2}u F_\pi^2(3-\gamma )tr[m]$ as in Eq.(\ref{SC}). The
dependence
of the $S^4$ coefficient on the explicit scale and chiral symmetry
breaking parameter $m$ is dictated in order for the symmetry to be
realized \`a la Nambu-Goldstone with $<0|S|0>=1$ and $<0|U|0>=I$ so that
$<0|D|0>=0$ and $<0|\pi^i|0>=0$. The vanishing of the $S^4$ coupling in
the chiral limit, $m \rightarrow 0$, is required since a potential of
the
form $\lambda S^4$ with $\lambda$ nonvanishing gives a classical
vacuum corresponding to $<S>_0 = 0$ which drives
$<D>_0 \rightarrow -\infty$.  This instability signals that the
corresponding effective Lagrangian realizes the symmetry in a 
Wigner-Weyl mode.  Consequently, a Nambu-Goldstone realization of the
symmetry requires a vanishing of the $S^4$ coupling in the chiral
limit, $m \rightarrow 0$. On the other hand, one cannot simply ignore
the $S^4$ term entirely since in its absence, the dilaton becomes
tachyonic for $m \neq 0$. In the exact symmetry chiral limit,
$m\rightarrow 0$, the invariant effective Lagrangian is simply obtained
as 
\be
{\cal{L}} = \frac{1}{2} F_D^2 \partial_\mu S \partial^\mu S 
                + \frac{1}{4} F_\pi^2 S^2 tr(\partial_\mu U^+ 
                  \partial^\mu U) .
\ee
   
Let us next consider nonlinear realizations of another type of
space-time symmetry, namely supersymmetry (SUSY). For spontaneously
broken SUSY, the dynamics of the Nambu-Goldstone fermion, the Goldstino,
is described by the Akulov-Volkov Lagrangian \cite{AV}. The nonlinear
SUSY transformations of the Goldstino fields are given by
\bea
\delta^Q(\xi, \bar\xi)\lambda^\alpha &=& F\xi^\alpha +\Lambda^\rho (\xi,
\bar\xi)\partial_\rho \lambda^\alpha \cr
\delta^Q(\xi, \bar\xi)\bar\lambda_{\dot\alpha} &=& F\bar\xi_{\dot\alpha}
+\Lambda^\rho (\xi, \bar\xi)\partial_\rho \bar\lambda_{\dot\alpha}  ,
\eea
where $\xi^\alpha$, $\bar\xi_{\dot\alpha}$ are the Weyl spinor SUSY
transformation parameters and $\Lambda^\rho (\xi, \bar\xi) \equiv
-\frac{i}{F}\left( \lambda \sigma^\rho \bar\xi -\xi\sigma^\rho
\bar\lambda \right)$ is a Goldstino field dependent translation vector
and $F$ is the Goldstino decay constant.  The Akulov-Volkov Lagrangian
takes the form
\be
{\cal L}_{AV} = -\frac{F^2}{2} \det{A} ,
\ee
with $A_\mu^{~\nu} =\delta_\mu^\nu
+\frac{i}{F^2}\lambda\stackrel{\leftrightarrow}{\partial}_\mu\sigma^\nu\bar{\lambda}$
the Akulov-Volkov vierbein.  Under the nonlinear SUSY variations, it
transforms as the total divergence
\be
\delta^Q(\xi,\bar{\xi}){\cal{L}}_{AV} = \partial_\rho \left(
\Lambda^\rho 
{\cal L}_{AV} \right)  ,
\ee
and hence the associated action $I_{AV} =\int d^4 x {\cal L}_{AV}$ is
SUSY invariant. 

Finally, let us explore the possibility of having simultaneous nonlinear
realizations of scale, chiral, and super symmetries. The superconformal
algebra \cite{WZ2}-\cite{F} nontrivially relates the SUSY and scale
symmetry transformations as 
\be
\left[ \delta^D (\epsilon), \delta^Q (\xi, \bar\xi) \right] =
\frac{1}{2}\delta^Q (\epsilon\xi, \epsilon\bar\xi) .
\ee
As a consequence, the Goldstino fields transform with scaling weight
$-\frac{1}{2}$ (recall the Nambu-Goldstone bosons of the spontaneously
broken global symmetry carry scaling weight $0$) as
\bea
\delta^D (\epsilon) \lambda_\alpha &=& \epsilon \left( -\frac{1}{2}
+x^\nu \partial_\nu \right) \lambda_\alpha \cr
\delta^D (\epsilon) \bar\lambda^{\bar\alpha} &=& \epsilon \left(
-\frac{1}{2} +x^\nu \partial_\nu \right) \bar\lambda^{\bar\alpha} .\cr
 & & 
\eea
As is the case for any matter field, the SUSY is nonlinearly realized
\cite{CL1}-\cite{CLLW} on the dilaton field as 
\be
\delta^Q (\xi, \bar\xi) D = \Lambda^\rho (\xi, \bar\xi) \partial_\rho D
.
\ee

We now attempt to construct an effective Lagrangian containing the
Goldstino, dilaton, and the Nambu-Goldstone bosons of the global
symmetry group $G$ which non-linearly realizes the scale, $G$, and super
symmetries, up to some soft explicit breakings of the scale and $G$
symmetries. To render the Akulov-Volkov Lagrangian scale invariant, we
simply multiply it by $S^4$ to raise its scaling weight to four. On the
other hand, the scale and $G$ invariant effective Lagrangian pieces can
be made nonlinearly SUSY invariant \cite{CL2}-\cite{CLLW} by simply
multiplying them by the Akulov-Volkov determinant, ${\rm det} A$, and
replacing all space-time derivatives by nonlinear SUSY covariant
derivatives so that, for example, $\partial_\mu \pi^i \rightarrow {\cal
D}_\mu \pi^i = (A^{-1})_\mu~^\nu\partial_\nu\pi^i$. So doing, we secure
the effective Lagrangian
\bea
{\cal L} &=& -\frac{F^2}{2} (\det{A})S^4 +\frac{F_D}{2} (\det{A}) {\cal
D}_\mu S {\cal D}^\mu S \cr
 & & +\frac{F_\pi^2}{4} (\det{A}) S^2 {\rm Tr}\left[ {\cal D}_\mu
U^\dagger {\cal D}^\mu U \right] +u F_\pi^2 (\det{A}) S^{3-\gamma} {\rm
Tr}\left[ m U^\dagger +U m \right] ,
\eea
where the last term is a soft explicit chiral and scale symmetry
breaking term. 

However, noting that ${\rm det} A$ starts with unity, then in order to
guarantee that the scale symmetry be realized in a Nambu-Goldstone
manner, the coefficient of the $S^4$ term which is the Goldstino decay
constant cannot be chosen  arbitrarily, but instead must be proportional
to the soft explicit chiral and scale symmetry breaking parameters.
Explicitly it is given by 
\be
F^2=(3-\gamma )u F^2_\pi {\rm Tr}[m]
\ee
The necessity of this identification is for exactly the same reason as
discussed in the non-SUSY case. Unless this coefficient vanishes in the
good symmetry chiral limit, the dilaton vacuum value is driven to
negative infinity. 

Of course, the fact that the Goldstino decay constant (and hence the
SUSY breaking scale) vanishes in the chiral limit raises its own host of
complications. First and foremost, it implies the scale symmetry and
SUSY cannot both be simultaneously realized as nonlinear symmetries.
Note that there is no difficulty in constructing an effective Lagrangian
invariant under both nonlinear SUSY and chiral symmetry transformations.
It is only when both SUSY and scale symmetry are to be realized
nonlinearly that one encounters the inconsistency. The source of the
problem can be directly traced to the fact that the Akulov-Volkov
Lagrangian relates the coefficients of the derivatively coupled
Goldstino self interactions and an overall constant which is the vacuum
energy accompanying the spontaneous superymmetry breaking. This constant
vacuum energy term is ignorable until one makes the model nonlinearly
scale invariant which is achieved by multiplying the entire
Akulov-Volkov Lagrangian by $S^4$. So doing, not only do the Goldstino
kinetic term and its self interactions get multiplied by this factor,
but so does the constant term. As such the erstwhile constant term now
becomes a dilaton self interaction potential term. In order for the
dilaton to be a Nambu-Goldstone particle, it cannot sustain a potential
whose coefficients are nonvanishing in the chiral limit. Thus the
coefficient of this term which is the Goldstino decay constant must
vanish in the chiral limit. But this is the same coefficient as that
multiplying the entire Akulov-Volkov determinant. Since the Goldstino
decay constant is the matrix element of the supersymmetry current
between the vacuum and the Goldstino state, its vanishing signals the
nonviability of the Goldstone realization. Consequently, both symmetries
cannot be simulataneously realized nonlinearly and the spectrum cannot
contain both a dilaton and a Goldstino as Nambu-Goldstone particles. We
have been unable to find a way out of this conundrum. One may try to
include additional low energy degrees of freedom and see if the
resulting model is consistent. Since $R$ symmetry is another component
of the superconformal algebra, we have investigated also including an
$R$-axion, the Nambu-Goldstone boson of spontaneous broken $R$ symmetry.
Once again the resultant effective Lagrangian does not have a consistent
interpretation in the chiral limit. It thus appears as if spontaneous
supersymmetry breaking requires the presence of explicit (hard) scale
symmetry breaking. This is certainly the case for the various models
which have been studied in the literature \cite{ADN}-\cite{NS}. Here we
have provided an argument that this is actually the case in general.
Note that the literature contains numerous studies of models  purporting
to include both dilatons and Goldstinos. In all of these models,
however, the dilaton aquires a non-zero vacuum expectation value.
Consequently, it is not really a Nambu-Goldstone boson and there is no
contradiction with our result. \\

\noindent
This work was supported in part by the U.S. Department 
of Energy under grant DE-FG02-91ER40681 (Task B).

\newpage
\end{document}